\documentstyle[12pt,amsfonts]{article}
\parskip=\medskipamount

\newcommand {\cD}{{\cal D}}

\newcommand {\cF}{{\cal F}}
\newcommand {\cG}{{\cal G}}
\newcommand {\cH}{{\cal H}}

\newcommand {\cK}{{\cal K}}
\newcommand {\cL}{{\cal L}}
\newcommand {\cM}{{\cal M}}

\newcommand {\cP}{{\cal P}}

\newcommand {\cV}{{\cal V}}

\newcommand {\cY}{{\cal Y}}

\newcommand{\bD}{{\bf D}}

\def\a{\alpha}
\def\b{\beta}

\def\d{\delta}

\def\f{\phi}

\def\G{\Gamma}

\def\l{\lambda}

\def\o{\omega}

\def\q{\theta}

\def\s{\sigma}
\def\t{\tau}

\def\x{\xi}
\def\z{\zeta}

\def\F{\Phi}
\def\J{\Psi}

\newcommand{\ad}{{\dot{\alpha}}}                           
\newcommand{\ve}{\varepsilon}                            
\newcommand{\cDB}{{\bar\cD}}                            

\newcommand{\pa}{\partial}                           

\newcommand{\sect}[1]{\setcounter{equation}{0}\section{#1}}

\newcommand{\be}{\begin{equation}}
\newcommand{\ee}{\end{equation}}
\newcommand{\bea}{\begin{eqnarray}}
\newcommand{\eea}{\end{eqnarray}}
\newcommand{\non}{\nonumber}

\begin{document}
\begin{titlepage}

\begin{flushright}
ITP-UH-15/97\\
hep-th/9705027 \\
\end{flushright}

\begin{center}
\large{{\bf The Higgs Mechanism in N = 2 Superspace} } \\
\vspace{1.0cm}

\large{Norbert Dragon and Sergei M. Kuzenko\footnote{ Alexander von 
Humboldt
Research Fellow. On leave from Department of Quantum Field Theory,
Tomsk State University, Tomsk 934050, Russia.}
} \\
\vspace{5mm}

\footnotesize{{\it Institut f\"ur Theoretische Physik, Universit\"at
Hannover\\
Appelstra{\ss}e 2, 30167 Hannover, Germany} \\
 }
\end{center}
\vspace{1.5cm}

\begin{abstract}
We describe the Higgs mechanism for general $N=2$ super Yang-Mills 
theories
in a manifestly supersymmetric form based on the harmonic superspace.
\end{abstract}
\vspace{15mm}

\vfill
\null
\end{titlepage}
\newpage
\setcounter{footnote}{0}

\sect{Introduction}

During the last two years, 
$N=2$, $D=4$ supersymmetric field theories have attracted
considerable interest kindled by the papers of Seiberg and 
Witten \cite{sw} on the exact determination of the low energy effective 
action in $N=2$ supersymmetric $SU(2)$ gauge models with spontaneously 
broken gauge symmetry (see Refs. \cite{bil,ler,agh} for a pedagogical
introduction).

The $N=2$ supersymmetric  gauge multiplet and its action \cite{n2} 
were constructed by Grimm, Sohnius and Wess
\cite{gsw,s} in  $N=2$ superspace ${\Bbb R}^{4|8}$ in terms of 
constrained superfields. In 
the harmonic superspace ${\Bbb R}^{4|8}\times S^2$
\cite{gikos,gios,zup} one can formulate the  $N=2$ supersymmetric
gauge models in terms of unconstrained (so-called analytic) superfields 
and study  quantum corrections in a manifestly supersymmetric fashion
\cite{gios,max,back}.

In this paper we investigate the spontaneous breakdown of gauge symmetry
for general $N=2$ supersymmetric gauge models in harmonic superspace
(see Ref. \cite{des} for a review of spontaneous symmetry breakdown in 
$N=1$ supersymmetric theories).
Such models admit two types of classical ground
states: the ones which are invariant under $SU(2)_A$, the (sub-) group of outer 
automorphisms of the  $N=2$ supersymmetry algebra,
and the  others which break  $SU(2)_A$.
We analyze in detail the first type of Higgs vacua. In this case only
vacuum expectation values of the scalar fields of the gauge multiplet and of
matter $\o$-hypermultiplets \cite{ddf} can occur. The theories possess 
three different physical phases which were described by Fayet \cite{fay2}
in the framework of $N=2$ supersymmetric grand unified theories.  

The paper is organized as follows. In section 2 we review the geometry of 
$N=2$ supersymmetric gauge models both in the standard $N=2$ superspace 
and in its harmonic extension. 
Section 3 is devoted to the analysis of the spontaneous breakdown of 
gauge symmetry for $N=2$ supersymmetric gauge models with matter. 
In section 4 we discuss the quantum equations of motion for the $N=2$
supersymmetric gauge multiplet.
The three appendices contain technical details. 

\sect{N = 2 super Yang-Mills geometry}

$N=2$ supersymmetric gauge models are constructed in 
$N=2$ superspace with coordinates
\be
z^M=(x^m,\theta_i^\alpha,
\bar \theta^i_{\dot{\alpha}}) \qquad \overline{\theta_i^\alpha}=\bar 
\theta^{{\dot{\alpha}}\,i} \qquad i=1,2
\ee
in terms of gauge covariant derivatives
\be
{\cal D}_M \equiv({\cal D}_m,{\cal D}^i_\alpha,
\bar {\cal D}^{\dot{\alpha}}_i)= D_M -
{\Bbb A}_M \qquad {\Bbb A}_M = A_M{}^a(z) \delta_a \qquad [\d_a,D_M] = 0
\label{1}
\ee
which are restricted by \cite{gsw,s}
\bea
\{{\cal D}^i_\alpha , \bar {\cal D}_{\dot{\alpha}j} \} &=&
-2{\rm i}\delta^i_j{\cal D}_{\alpha\dot{\alpha}} \nonumber \\
 \{{\cal D}^i_\alpha ,{\cal D}^j_\beta \} =
-2
\varepsilon_{\alpha\beta}\varepsilon^{ij} \bar {\Bbb W}   \quad &{}& 
\quad 
\{\bar {\cal D}_{\dot{\alpha}i} ,\bar {\cal D}_{\dot{\beta}j}\}
= -2
\varepsilon_{\dot{\alpha}\dot{\beta}}\varepsilon_{ij}{\Bbb W}
\label{2} \\
\left [ {\cal D}_{\alpha\dot{\alpha}}, {\cal D}^i_{\beta} \right ] =
{\rm i}
\varepsilon_{\alpha\beta}(\cDB^i_{\dot{\alpha}}
\bar {\Bbb W})   \quad &{}& \quad
\left [ {\cal D}_{\alpha \dot{\alpha}} ,\bar {\cal D}_{\dot{\beta}i}
\right ] = 
{\rm i}
\varepsilon_{\dot{\alpha}\dot{\beta}} (\cD_{\alpha i} {\Bbb W})\;. 
\nonumber
\end{eqnarray}
Here $\Bbb W$ is a complex linear combination of 
generators $\delta_a$ of a real Lie algebra
\bea
&{\Bbb W}=W^a(z)\delta_a
\qquad \bar {\Bbb W} = \bar{W}^a(z)\delta_a \label{w}\\
& \left [ \delta_a , \delta_b \right ] = f_{ab}{}^c \delta_c
\qquad [\d_a , \cD_M] = 0\ . \label{alg}
\eea
If one applies $\delta_a$ to tensors $\phi$ one obtains
$\delta_a \phi = - {\rm i}T_a \phi$ with matrices $T_a$ which 
are typically taken to be
hermitean, $(T_a)^\dag = T_a$, 
such that  ${\rm i}T_a$ generate a unitary representation 
of the gauge group and satisfy the same Lie algebra as $\delta_a$. 
It is, however, worthwhile
to allow for more general $\d_a$ even 
though we will not pursue such a generalization 
here. Below we will also make use of the notation
\be
A_M = A_M{}^a T_a \qquad W = W^a T_a \qquad \bar W = {\bar W}^a T_a 
= W^\dag 
\label{notation}
\ee 
for unitary representations of the gauge group.

The super field strengths $W^a$ satisfy the Bianchi identities
\be
\cDB_{\ad i} {\Bbb W} = 0 \qquad \cD^{\a(i} \cD_\a^{j)} {\Bbb W} 
= \cDB^{(i}_\ad 
\cDB^{j)\ad}
\bar {\Bbb W} \;.
\label{3}
\ee
The transformation laws of $\cD_M$ and of matter multiplets $U(z)$ read
\be
\cD'_M = {\rm e}^{\t} \,\cD_M \,{\rm e}^{-\t} 
\qquad U'={\rm e}^{\t}\, U \qquad \t = \t^a(z) \d_a
\label{4}
\ee
where the gauge parameters $\t^a$ are unconstrained real superfields.  

An important feature of the $N=2$ supersymmetric gauge multiplet, 
in contrast to
the  $N=1$ case, is that one can have a covariantly constant 
super field strength 
\be
\cD^i_\a {\Bbb W} = 0 \quad \Rightarrow \quad 
[\bar {\Bbb W}, {\Bbb W}] =0\qquad \cD_m {\Bbb W} = 0\;.
\label{5}
\ee
With the gauge transformations (\ref{4}) one can cast 
the background value of the
covariant derivatives into the form \cite{gsw}
\be
\bD^i_\a =D^i_\a - 
\q^i_\a \bar {\Bbb W}_0 \qquad
\bar{\bD}_{\ad i} =\bar D_{\ad i}+ 
\bar{\q}_{\ad i} {\Bbb W}_0 \qquad 
\bD_m =\pa_m
\label{6}
\ee
with
\be
[\bar {\Bbb W}_0, {\Bbb W}_0] = 0 \qquad 
{\Bbb W}_0 =W^a_0\delta_a \qquad W^a_0= {\rm const}\;.
\label{7}
\ee
Such a gauge fixing is super Poincar\'e covariant
provided every supersymmetry transformation
\be
\delta U = (\epsilon^\alpha_i Q^i_\alpha +
\bar \epsilon_{\dot{\alpha}}^i \bar Q^{\dot{\alpha}}_i)U
\label{8}
\ee
is accompanied by the $\epsilon$-dependent gauge
transformation
\be
\delta U =  \tau \, U \qquad
\tau = \epsilon^\alpha_i \theta^i_\alpha \bar {\Bbb W}_0 -
\bar \epsilon_{\dot{\alpha}}^i {\bar \theta}^{\dot{\alpha}}_i {\Bbb W}_0 \;.
\label{9}
\ee
As a result, eq. (\ref{8}) turns into
\be
\delta U = (\epsilon {\bf Q} +
\bar \epsilon \bar {\bf Q})U
\label{10}
\ee
where
\be
{\bf Q}^i_\alpha = \frac{\partial}{\partial
\theta^\alpha_i} - {\rm i}\bar \theta^{\dot{\alpha}i}
\partial_{\alpha \dot{\alpha}}
+ \theta^i_\alpha \bar {\Bbb W}_0 
\qquad
\bar {\bf Q}_{\dot{\alpha} i} = - \frac{\partial}
{\partial \bar \theta^{\dot{\alpha}i}} + {\rm i}  \theta^{\alpha}_i
\partial_{\alpha \dot{\alpha}}
- \bar \theta_{\dot{\alpha}i} {\Bbb W}_0\;.
\label{11}
\ee
These generators form the $N=2$ supersymmetry
algebra with central charges ${\Bbb W}_0$ and ${\bar {\Bbb W}}_0$
\bea
\{ {\bf Q}^{i}_{\alpha} ,\bar {\bf Q}_{\dot{\alpha}j}\} &=&
2 {\rm i}\delta^i_j \partial_{\alpha\dot{\alpha}} \nonumber\\
\{ {\bf Q}^i_\alpha ,{\bf Q}^j_\beta\} =
2 \varepsilon_{\alpha\beta}\varepsilon^{ij} \bar {\Bbb W}_0\quad &{}& 
\quad
\{ \bar {\bf Q}_{\dot{\alpha}i}
\bar {\bf Q}_{\dot{\beta}j} \} =
2 \varepsilon_{\dot{\alpha}\dot{\beta}}
\varepsilon_{ij} {\Bbb W}_0
\label{12}
\eea
and eq. (\ref{6}) defines the corresponding covariant derivatives,
$\{{\bf Q}^i_\alpha , {\bf D}^j_\beta \} =
\{{\bf Q}^i_\alpha , \bar {\bf D}_{\dot{\beta}j}\}
=0$. Because $(\bar W_0 W_0) = |Z|^2 \le M^2$ in unitary representations, a 
constant  background of the scalar field of the supersymmetric gauge multiplet 
generates masses for matter multiplets.

In the harmonic superspace ${\Bbb R}^{4|8} \times S^2$
one  can solve \cite{gikos,gios,zup}  the constraints (\ref{2}) and (\ref{3})
in terms of unconstrained analytic superfields
and derive manifestly supersymmetric Feynman rules.

It is useful to parameterize the two-sphere $S^2 =SU(2)/U(1)$ by 
harmonics, i.e.
group elements 
\bea
&({u_i}^-\,,\,{u_i}^+) \in SU(2)\non\\ 
&u^+_i = \ve_{ij}u^{+j} \qquad \overline{u^{+i}} = u^-_i 
\qquad u^{+i}u_i^- = 1 \;.
\eea
Then tensor fields over $S^2$ are in a 
one-to-one correspondence with functions over $SU(2)$ of definite
$U(1)$-charges. A function $\Psi^{(p)}(u)$ is said to have $U(1)$-charge 
$p$
if 
$$
\Psi^{(p)}({\rm e}^{{\rm i}\a} u^+,{\rm e}^{-{\rm i}\a} u^-)=
{\rm e}^{{\rm i}\a p} \Psi^{(p)}(u^+,u^-) \qquad |{\rm e}^{{\rm i}\a}| = 
1\;.
$$ 
The operators
\bea
& D^{\pm \pm} =u^{\pm i} \frac{\pa}{\pa u^{\mp i}} \qquad
D^0 = u^{+i} \frac{\pa}{\pa u^{+i}} - u^{-i} \frac{\pa}{\pa u^{-i}} \non 
\\
& [D^0,D^{\pm \pm}] = \pm 2D^{\pm \pm} \qquad [D^{++}, D^{--}] = D^0
\label{13}
\eea
are left-invariant vector fields on $SU(2)$ and $D^0$ is the 
$U(1)$-charge operator.

With use of the harmonics one can convert the spinor covariant 
derivatives
into $SU(2)$-invariant operators on ${\Bbb R}^{4|8} \times S^2$
\be
{\cal D}^\pm_\alpha = {\cD}^i_\alpha u^\pm_i
\qquad {\bar{\cal D}}^\pm_{\dot\alpha}={\bar{\cD}}^i_{\dot\alpha} u^\pm_i 
\;.
\label{14}
\ee
Then it follows from (\ref{2})
\be
\{\cD^+_\a,\cD^+_\b\} = \{\cDB^+_ {\dot\a}, 
\cDB^+_{\dot\b}\}= \{\cD^+_\a, \cDB^+_{\dot\a} \}=0
\label{15}
\ee
which can be solved by
\be
\cD^+_\a = {\rm e}^{{\Bbb G}}\,D^+_\a \,{\rm e}^{-{\Bbb G}} \qquad
\cDB^+_{\dot\a}={\rm e}^{{\Bbb G}}\,{\bar D}^+_{\dot\a}\,
{\rm e}^{-{\Bbb G}} \qquad {\Bbb G} = G^a(z,u)\d_a \;.
\label{16}
\ee
Here the superfields $G^a$ have vanishing 
$U(1)$-charge and are real, $\breve{G}^a = G^a$, with respect to the 
analyticity preserving conjugation 
$\; \breve{} \;\equiv \;\stackrel{\star}{\bar{}}$ \cite{gikos}, where
the operation ${}^\star$ is defined by $(u^+_i)^\star = u^-_i$, 
$(u^-_i)^\star = - u^+_i$, hence $(u^{\pm}_i)^{\star \star} = - u^{\pm}_i$. 
Eq. (\ref{16}) 
partially solves the constraints (\ref{2}). An obvious consequence of
the relations (\ref{15}) and (\ref{16}) is that the harmonic superspace
allows one to introduce new superfield types, i.e. covariantly analytic
superfields constrained by
\be
\cD^+_\a\F^{(p)}=\cDB^+_{\dot\a} \F^{(p)}=0
\label{17}
\ee
and hence
\be
\F^{(p)}={\rm e}^{{\Bbb G}}\,\f^{(p)} \qquad
D^+_\a \f^{(p)}={\bar D}^+_{\dot\a} \f^{(p)}=0\;.
\label{18}
\ee
The superfield $\f^{(p)}$ turns out to be unconstrained  
over an analytic
subspace of the harmonic superspace parameterized by $\z^M_A\equiv\{
x^m_A,\q^{+\a},{\bar\q}^+_{\dot\a}\}$ and $u^\pm_i$, where \cite{gikos} 
$$
x^m_A = x^m - 2{\rm i} \q^{(i}\s^m {\bar \q}^{j)}u^+_i u^-_j \qquad
\q^\pm_\a=u^\pm_i \q^i_\a \qquad {\bar \q}^\pm_{\dot\a}=u^\pm_i{\bar
\q}^i_{\dot\a} \;.
$$ 
Another crucial
consequence of (\ref{15}) and (\ref{16}) is that the gauge group for the
prepotential $\Bbb G$ is larger than the original $\t$-group (\ref{4}):
\be
{\rm e}^{{\Bbb G}'} = {\rm e}^{\t} \,{\rm e}^{{\Bbb G}} \,
{\rm e}^{-\l} \qquad \l=\l^a(\z_A,u)\d_a \qquad 
D^+_\a \l^a = {\bar D}^+_\ad \l^a=0 \;.
\label{19}
\ee
Here the unconstrained analytic gauge parameters $\l^a$ have vanishing 
$U(1)$-charge. They are real, 
$\breve{\l}^a = \l^a$, with
respect to the analyticity preserving conjugation. The set of all 
$\l$-transformations is called the $\l$-group. The $\t$-group acts 
on
$\Phi^{(p)}$ and leaves $\f^{(p)}$ unchanged; the $\l$-group acts on
$\f^{(p)}$ by 
\be
\phi'^{(p)}={\rm e}^{\l} \f^{(p)}
\label{19-1}
\ee
and leaves $\Phi^{(p)}$ unchanged. The $\F^{(p)}$ and $\f^{(p)}$
describe the covariantly analytic superfield in $\t$-
and $\l$-frame respectively.

In the $\t$-frame, the complete set of gauge-covariant derivatives reads
\be
\cD_{\underline{M}}\equiv(\cD_M,\cD^{++},\cD^{--},
\cD^0) \qquad \cD^{\pm\pm}=D^{\pm\pm} \qquad
\cD^0=D^0
\label{20}
\ee
and their transformation law is the same as that of $\cD_M$ given by
(\ref{4}). In the $\l$-frame, the covariant derivatives 
\be
\nabla_{\underline{M}}={\rm e}^{-{\Bbb G}}\,{\cD}_{\underline{M}}\,
{\rm e}^{{\Bbb G}}
\label{21}
\ee
transform by the rule
\be 
\nabla'_{\underline{M}} =  {\rm e}^{\l} \,\nabla_{\underline{M}}\,
{\rm e}^{-\l}
\label{22}
\ee
and their algebra reads
\bea
&{}& \qquad \qquad \{{\bar \nabla}^+_{\dot\a},\nabla^-_\a\}=-\{\nabla^+_
\a,{\bar \nabla}^-_{\dot\a}\}=2{\rm i}\nabla_{\a{\dot\a}} \label{23}\\
&{}&\{\nabla^+_\a,\nabla^-_\b\}=2\ve_{\a\b}
{\bar {\Bbb W}}_\t \qquad  \quad
\{{\bar \nabla}^+_{\dot\a},{\bar \nabla}^-_{\dot\b}\}=-2
\ve_{{\dot\a}{\dot\b}}{\Bbb W}_\t  \label{24}\\
&{}& [\nabla^{\pm \pm},\nabla^\mp_\a]=\nabla^\pm_\a \qquad \qquad \qquad
[\nabla^{\pm \pm},{\bar \nabla}^\mp_{\dot\a}]={\bar \nabla}^\pm_{\dot\a}
\label{25}\\
&{}&[\nabla^0, \nabla^{\pm\pm}] =\pm 2 \nabla^{\pm\pm} \qquad \qquad
[\nabla^{++},\nabla^{--}]= \nabla^0
\label{26}
\eea
where
\be
{\Bbb W}_\l = {\rm e}^{-{\Bbb G}}\, {\Bbb W} \,{\rm e}^{{\Bbb G}}\qquad
{\bar {\Bbb W}}_\l = {\rm e}^{-{\Bbb G}}\, {\bar {\Bbb W}} \, 
{\rm e}^{{\Bbb G}}\;.
\label{27}
\ee
The other (anti-)commutators vanish except those involving vector
covariant derivatives, the latter can be readily obtained from
the relations given.

In the $\l$-frame, we have
\bea
&{}& \nabla^+_\a=D^+_\a \qquad {\bar\nabla}^+_{\dot\a}=
{\bar D}^+_{\dot\a} \qquad \nabla^0 = D^0 \non \\
&{}& \nabla^{\pm\pm}= {\rm e}^{-{\Bbb G}}\,D^{\pm\pm} \,{\rm e}^{{\Bbb G}}
=D^{\pm \pm} - {\Bbb V}^{\pm \pm} \qquad {\Bbb V}^{\pm \pm} 
=V^{\pm \pm a}\d_a\;. 
\label{28}
\eea
Since $[\nabla^{++},\nabla^+_\a] = [\nabla^{++},{\bar \nabla}^+_\ad] = 
0$,
the connection components $V^{++a}$ prove to be analytic real 
superfields,
$D^+_\a V^{++a} = {\bar D}^+_\ad V^{++a} =0$,
$\breve{V}^{++a}=V^{++a}$, with the transformation law
\be
{\Bbb V}'^{++}={\rm e}^{\lambda}\,{\Bbb V}^{++}\,{\rm e}^{-\lambda}-
{\rm e}^{\lambda}\,D^{++}\,{\rm e}^{-\lambda}\;. 
\label{29}
\ee
Using the (anti-)commutation relations for the covariant derivatives, 
especially eq. (\ref{25}), as well as the explicit form (\ref{28})
of $\nabla^+_\a$ and ${\bar \nabla}^+_\ad$, one easily expresses the
connections associated with $\nabla_M$ in terms of ${\Bbb V}^{--}$. In 
particular,
the super field strengths read
\be
{\Bbb W}_\l = \frac{1}{4} (\bar D^+)^2 {\Bbb V}^{--} \qquad
\bar {\Bbb W}_\l = \frac{1}{4} (D^+)^2 {\Bbb V}^{--}\;.
\label{30}
\ee
${}$This implies
\be
(\nabla^+)^2 {\Bbb W}_\l = ({\bar \nabla}^+)^2 \bar {\Bbb W}_\l \;.
\label{31}
\ee
The remaining Bianchi identities 
\bea
(\nabla^-\nabla^+ +\nabla^+ \nabla^-){\Bbb W}_\l &=& 
({\bar \nabla}^-{\bar\nabla}^+ + {\bar \nabla}^+{\bar\nabla}^-) 
{\bar {\Bbb W}}_\l \non \\
(\nabla^-)^2 {\Bbb W}_\l &=& ({\bar \nabla}^-)^2 {\bar {\Bbb W}}_\l
\label{32}
\eea 
are trivial consequences of the covariant $u$-independence 
of ${\Bbb W}_\l$ and
${\bar {\Bbb W}}_\l$ ($\nabla^{\pm \pm}{\Bbb W}_\l 
=\nabla^{\pm \pm}{\bar {\Bbb W}}_\l = 0$)
and of the identities
$$
[\nabla^{--},(\nabla^+)^2] = \nabla^-\nabla^+ + \nabla^+\nabla^-\qquad
[\nabla^{--},[\nabla^{--},(\nabla^+)^2]] = 2(\nabla^-)^2
$$
and their analogs with $\nabla^+$'s replaced by ${\bar \nabla}^+$'s.
Further, the relation $[\nabla^{++},\nabla^{--}] = D^0$ can be treated
as an equation uniquely determining ${\Bbb V}^{--}$ in terms of 
${\Bbb V}^{++}$. It is given by \cite{zup}
\be
{\Bbb V}^{--}(z,u)=
-\sum\limits_{n=1}^\infty 
\int du_1 du_2\dots du_n\frac{{\Bbb V}^{++}(z,u_1){\Bbb V}^{++}(z,u_2)
\cdots {\Bbb V}^{++}(z,u_n)}
{(u^+u^+_1)(u^+_1u^+_2)\cdots(u^+_nu^+)}
\label{33}
\ee
where the integration over $SU(2)$ is defined by \cite{gikos}
$$
\int du \; 1 = 1 \qquad \int du \, u^+_{(i_1} \ldots u^+_{i_n} u^-_{j_1}
\ldots u^-_{j_m)} = 0 \qquad n+m > 0
$$
and the properties of harmonic
distributions are described in \cite{gios}.
As a result, all geometric objects are expressed in terms of the single
unconstrained analytic real prepotential ${\Bbb V}^{++}$.
              
In the next sections, we will restrict ourselves by the study of
unitary matrix representations of the gauge group and make use of the
notation
\be
G = G^a T_a \qquad V^{\pm \pm} = V^{\pm \pm a} T_a \;.
\ee

The gauge freedom (\ref{29}) can be used to choose the Wess-Zumino
gauge \cite{gikos}
\bea
&{}&V^{++}(x_A, \q^+, {\bar \q}^+, u)
= \q^+ \q^+{\bar N}(x_A)
+ {\bar \q}^+ {\bar \q}^+ N(x_A) \non\\
&{}&-2 {\rm i}\q^+ \s^m {\bar \q}^+ V_m(x_A) 
+ {\bar \q}^+{\bar \q}^+\q^{+\a} \J^i_\a (x_A) u^-_i
+ \q^+ \q^+{\bar \q}^+_\ad {\bar \J}^{i_\ad}(x_A) u^-_i\non\\ 
&{}&+ {\rm i}\,\q^+ \q^+{\bar \q}^+ {\bar \q}^+ D^{(ij)}(x_A) u^-_i  
u^-_j 
\label{34}
\eea
where the triplet $D^{ij}$ satisfies the reality condition (\ref{rc}).
Thus we stay with the field multiplet of $N=2$ supersymmetric gauge theory \cite{n2}.
The residual gauge freedom is given by $\l^a = \x^a(x_A)$ describing the 
standard Yang-Mills transformations.

In the case of constant curvature, the prepotentials
$\Bbb G$ and ${\Bbb V}^{\pm \pm}$ read  
\be
{\Bbb G}_0 = \q^{- \a}\q^+_\a {\bar {\Bbb W}}_0 +
{\bar \q}^{-}_{\ad}{\bar \q}^{+\ad} {\Bbb W}_0
\qquad
{\Bbb V}_0^{\pm \pm}= -\q^{\pm \a}\q^\pm_\a {\bar {\Bbb W}}_0 -
{\bar \q}^{\pm}_\ad {\bar \q}^{\pm \ad} {\Bbb W}_0\;.
\label{35}
\ee

\sect{Spontaneous breakdown of gauge symmetry}

We consider a general $N=2$ supersymmetric gauge theory with 
matter hypermultiplets being
described by unconstrained analytic superfields 
$\{q^+(\z_A,u), \breve{q}^+(\z_A,u)\}$ ($q$-hypermultiplet) and 
$\o(\z_A,u)$,
$\breve{\o} = \o$
(real $\o$-hypermultiplet) \cite{gikos} in some representations of the 
gauge
group. The gauge-invariant action is given by 
\be 
S[V^{++}, q^+, \o] = S_{\rm SYM}[V^{++}] +S_{\rm MAT}[V^{++}, q^+, \o]
\label{3.1}
\ee
where the pure $N=2$ supersymmetric gauge action has 
the form \cite{gsw,gikos,zup}
\bea
&{}& S_{{\rm SYM}}[V^{++}]=\frac{1}{2g^2} {\rm tr}\int d^4xd^4\theta\, 
W^2=
\frac{1}{2g^2}{\rm tr} \int d^4xd^4{\bar \theta}\, {\bar W}^2 
\label{3.2}\\
&{}& =\frac{1}{g^2} {\rm tr}\,
\int d^{12}z\sum\limits_{n=2}^\infty\frac{(-{\rm i})^n}
{n}\int du_1 du_2\dots du_n\frac{V^{++}(z,u_1)V^{++}(z,u_2)
\cdots V^{++}(z,u_n)}
{(u^+_1u^+_2)(u^+_2u^+_3)\cdots(u^+_nu^+_1)}
\non
\eea                                   
with ${\rm tr}\, (T_aT_a) = \d_{ab}$. The matter action reads
\cite{gikos}
\be
S_{{\rm MAT}}[V^{++}, q^+, \o]=
- \int d\z^{(-4)}
du \,\breve{q}^+ \nabla^{++}q^+ -
\frac{1}{2}\int d\zeta^{(-4)}
du\,\nabla^{++}\omega^{\rm T}\nabla^{++}\omega
\label{3.3}
\ee
with $d\z^{(-4)}=d^4x_A d^2\q^+d^2{\bar \q}^+$. It is not difficult to 
derive the dynamical equation for $V^{++}$ (see Appendix C):
\be 
\frac{1}{4g^2} (\nabla^+)^2 W^a_\l - {\rm i}\breve{q}^+ T_a q^+ 
+{\rm i}\o^{\rm T} T_a \nabla^{++} \o  =0\;.
\label{3.4}
\ee
Therefore, the theory possesses $SU(2)_A$-invariant solutions of the form
\be
\cD^i_\a W=0 \qquad q^+= 0 \qquad \nabla^{++}\o =0 
\label{3.5}
\ee
which correspond to the possible $SU(2)_A$-invariant Higgs vacua.
The importance of $\o$-hypermultiplets for realizing $SU(2)_A$-invariant 
Higgs
vacua was first recognized by Delamotte, Delduc and Fayet \cite{ddf}.

More generally, there may exist vacuum solutions with broken
$SU(2)_A$. Such solutions
are described by the requirements
\bea
\cD^i_\a W=0 \quad && \quad 
\breve{q}^+ T_a q^+ = \o^{\rm T} T_a \nabla^{++} \o \non \\
\nabla^{++} q^+= 0 \quad && \quad 
\nabla^i_\a q^+={\bar \nabla}_{\ad i} q^+= 0 \non \\
(\nabla^{++})^2 \o = 0 \quad && \quad  
\nabla^i_\a \o ={\bar \nabla}_{\ad i} \o = 0\;.
\label{3.5-1}
\eea
In what follows, we restrict ourselves by the study of the 
$SU(2)_A$-invariant Higgs vacua. 

The above solutions are restricted by some consistency conditions.
First of all, the requirement of $W$ to be covariantly constant
implies $[{\bar W}, W] = 0$. Another consistency 
condition follows from the fact that for negative $p$
the equation $D^{++}f^{(p)}(u) = 0$
has the unique solution $f^{(p)}= 0$. Therefore, if we have a 
scalar superfield $\f (z,u)$ constrained by $D^{++}\f= 0$, then 
$D^{--}\f= 0$ also holds ($D^{--}D^{++}\f = D^{++}D^{--}\f = 0$), 
and hence
$\f$ is $u$-independent, $\f = \f(z)$. If, in addition, $\f$ is an 
analytic superfield by construction, then we automatically have 
$\f = {\rm const}$
since the analyticity requirements 
$D^+_\a \f = u^+_i D^i_\a \f(z) = 0$ and 
${\bar D}^+_\ad \f = u^+_i {\bar D}^i_\ad \f(z) = 0$ are now equivalent
to $D^i_\a \f = {\bar D}^i_\ad \f = 0$. Keeping all this in mind,
we analyze the last equation in (\ref{3.5}). By definition
\be
\nabla^{++} \o ={\rm e}^{{\rm i} G}D^{++} {\rm e}^{-{\rm i} G} \o = 0
\quad \Rightarrow \quad \o = {\rm e}^{{\rm i} G} \o_\t (z)\;.
\label{3.6}
\ee
Therefore, $\o$ is $u$-independent in the $\t$-frame. Then, however,
the analyticity requirements imply  
\be
\cD^i_\a \,\o_\t = 0 \qquad {\bar \cD}^i_\ad \,\o_\t = 0\;.
\label{3.7}
\ee
These are consistent only if 
\be
{\bar W}\, \o_\t = 0 \qquad W \,\o_\t = 0
\label{3.8}
\ee
and then $\o_\t \equiv \o_0 = {\rm const}$. Let us choose the gauge in 
which 
$W = W_0 = {\rm const}$. The explicit form of ${\Bbb G}_0$ (\ref{35})
and eq. (\ref{3.6}) tell us that $\o = \o_0$. In summary, the admissible
$SU(2)$-invariant
Higgs vacua are parameterized by the expectation values $W_0$, ${\bar 
W}_0$
and $\o_0$ constrained by
\be
[{\bar W}_0, W_0] = 0 \qquad W_0\,\o_0 = {\bar W}_0\,\o_0 = 0\;.
\label{3.9}
\ee
Physically, the three choices
(i) $W_0 \neq 0$, $\o_0 = 0$; (ii) $W_0 \neq 0$, $\o_0 \neq 0$;
(iii) $W_0 = 0$, $\o_0 \neq 0$ describe different phases of the theory.

Let us first examine the case $W_0 \neq 0$, $\o_0 = 0$.
We choose the supersymmetric gauge in which the vacuum covariant 
derivatives look like in eq. (\ref{6}). It is supersymmetric since
the Higgs vacuum conditions are invariant under the gauge transformation
with parameter (\ref{9}). Then we still have an unbroken
gauge group generated by the subalgebra $\cY$ of elements of the Lie 
algebra
$\cG$ which commute with 
$({\rm Re}\, W_0^a)\,T_a$ and $({\rm Im}\, W_0^a)\,T_a$. As is obvious, 
$\cY$
includes the abelian subalgebra $\cH$ spanned by 
$({\rm Re}\, W_0^a)\,T_a$ and $({\rm Im}\, W_0^a)\,T_a$.
But now, however, we deal with supersymmetry with central charges, 
and there appear massive superfields:
not only several $q$- and $\o$-hypermultiplets, but also the 
components of the gauge multiplet $\{V^{++a}\}$ which belong 
to the orthogonal complement $\cK$ to $\cY$ 
in the Lie algebra $\cG$ of the gauge group, $\cG = \cY \oplus \cK$.      
Herewith all the massive superfields, not only the massive 
hypermultiplets,
describe short representations of the $N=2$ supersymmetry with central 
charges,
since the mass matrix turns out to look like 
\be 
M^2 = W_0{\bar W}_0={\bar W}_0 W_0
\label{3.10}
\ee
and, hence, the values of the mass and central charge coincide, 
as a consequence of (\ref{12}).

The simplest way to prove the above assertion is to analyze
the gauge structure upon the spontaneous breakdown. Let us represent 
the gauge superfield in the form
\be 
V^{++} = V^{++}_0 + \cV^{++}
\label{3.11}
\ee
where $V^{++}_0$, given by (\ref{35}), corresponds to the Higgs vacuum
and $\cV^{++}$ describe deviations from the ground state. 
The gauge transformation (\ref{29}) turns into 
\be
\d \cV^{++} = D^{++}\l + {\rm i}[V^{++}_0,\l] + {\rm i}[\cV^{++},\l] 
=\bD^{++}\l + {\rm i} [\cV^{++},\l] \qquad \l = \l^a T_a \;.
\label{3.12}
\ee
As in the case of unbroken gauge symmetry we can impose the 
Wess-Zumino gauge. We can use the residual gauge transformations
with real parameters $\x^a$
\be 
\d N = {\rm i} [N , \x] -{\rm i} [W_0,\x] \qquad \x = \x^a T_a
\label{3.13}
\ee
to gauge away a half of the complex $N$'s which
correspond to the broken symmetries. The spin content of the multiplets
obtained is $2\,({\bf 0} \oplus \frac{{\bf 1}}{{\bf 2}} \oplus 
\frac{{\bf 1}}{{\bf 2}} \oplus {\bf 1})$ and the doubling of fields
is caused by the central charges $W_0$ and ${\bar W}_0$
in the supersymmetry algebra. This
corresponds to the short massive multiplet with highest spin one
\cite{fay}.

To analyze the mass spectrum of the theory, we insert
(\ref{3.11}) in the action functional (\ref{3.1}). This gives for the 
supersymmetric gauge action
\be
S_{{\rm SYM}}=
\frac{1}{g^2} {\rm tr}\,
\int d^{12}z\sum\limits_{n=2}^\infty\frac{(-{\rm i})^n}
{n}\int du_1 du_2\dots du_n\frac{\cV_\t^{++}(z,u_1)\cV_\t^{++}(z,u_2)
\cdots V_\t^{++}(z,u_n)}
{(u^+_1u^+_2)(u^+_2u^+_3)\cdots(u^+_nu^+_1)}
\label{3.14}
\ee
where
\be
\cV_\t^{++} ={\rm e}^{-{\rm i}G_0}\cV^{++}{\rm e}^{{\rm 
i}G_0}\;.
\label{3.15}
\ee
The matter action takes the form
\be
S_{{\rm MAT}}=
-\int d\z^{(-4)}
du \,\breve{q}^+ (\bD^{++} + {\rm i}\cV^{++})q^+ 
+ \frac{1}{2}\int d\zeta^{(-4)}
\o^{\rm T} (\bD^{++} + {\rm i}\cV^{++})^2\o \;.
\label{3.16}
\ee
Both $S_{{\rm SYM}}$ and $S_{{\rm MAT}}$ are manifestly invariant under
the supersymmetry transformations (\ref{10}) generated by (\ref{11}).
They also are invariant under 
the gauge transformations (\ref{3.8}) supplemented
by those of the matter superfields (\ref{19-1}). The gauge freedom can be
fixed by imposing the gauge condition 
\be
\bD^{++} \cV^{++} = 0
\label{gk}
\ee
or, equivalently, by adding to $S_{{\rm SYM}}$ the following gauge-fixing
term (see Refs. \cite{gios,back} for more details)
\bea
S_{{\rm GF}}[\cV^{++}]=\frac{1}{2g^2\a}{\rm tr}\,
\int d^{12}zdu_1du_2\frac
{(u^-_1u^-_2)}{(u^+_1u^+_2)^3}(D^{++}_1\cV^{++}_\tau(1))(D^{++}_2
\cV^{++}_\tau(2))\non \\
=\frac{1}{2g^2\a}{\rm tr}\,\int d^{12}zdu_1du_2\,\frac
{\cV^{++}_\tau(1)\cV^{++}_\tau(2)}{(u^+_1u^+_2)^2}-\frac{1}{4g^2\alpha}
{\rm tr}\,\int
d^{12}zdu\,\cV^{++}_\tau(D^{--})^2\cV^{++}_\tau\;.
\label{3.17}
\eea
The equations of motion corresponding to $S_{{\rm SYM}} + S_{{\rm GF}}$
should be supplemented by the gauge condition (\ref{gk}).

For the special choice $\a = -1$ we obtain
\bea
&{}&S_{{\rm SYM}} + S_{{\rm GF}} 
= -\frac{1}{2g^2} {\rm tr}\,\int d\zeta^{(-4)}du\,
\cV^{++}\stackrel{\sim}{\Box}\cV^{++} \non\\
&{}&+ \frac{1}{g^2}\,{\rm tr}\,\int d^{12}zdu_1\dots du_n
\sum\limits_{n=3}^\infty
{\frac{(-i)}{n}}^{n}\frac{\cV^{++}_\tau(z,u_1)\cdots
\cV^{++}_\tau(z,u_n)}{(u^+_1u^+_2)\dots(u^+_nu^+_1)}\;. 
\label{3.18}
\eea
Here we have used the relation\footnote{We use the notation
$(D^+)^4 = \frac{1}{16} (D^+)^2 ({\bar D}^+)^2$,
$(D^\pm)^2=D^{\pm \alpha} D^\pm_\alpha$,
$({\bar D}^\pm)^2 = {\bar D}^\pm_{\dot{\alpha}}{\bar D}^{\pm 
\dot{\alpha}}$
and similar notation for the gauge-covariant derivatives.}
\bea
\frac{1}{2}{\rm tr}\int
d^{12}zdu\,\cV^{++}_\tau(D^{--})^2\cV^{++}_\tau
&=&\frac{1}{2}{\rm tr}\,\int                                           
d^{12}zdu\,\cV^{++}(\bD^{--})^2\cV^{++}\non\\
&=& - {\rm tr}\,\int d\zeta^{(-4)}du\,
\cV^{++}\stackrel{\sim}{\Box}\cV^{++} 
\label{3.19}
\eea
where
\be
\stackrel{\sim}{\Box} = -\frac{1}{2} (D^+)^4 (\bD^{--})^2 \qquad
\stackrel{\sim}{\Box} \f^{(p)} = (\Box + {\bar W}_0 W_0)\f^{(p)} 
\label{3.20}
\ee
for any analytic superfield $\f^{(p)}$ in arbitrary representation of the 
gauge group.

${}$From eqs. (\ref{3.14}) and (\ref{3.16}) we can single out 
the part which is quadratic in
the superfields  
\be
S_{(2)}=
\int d\z^{(-4)} du \, \left\{
-\frac{1}{2g^2} {\rm tr}\,\cV^{++}\stackrel{\sim}{\Box}\cV^{++}  
-\breve{q}^+\bD^{++}q^+
+\frac{1}{2} \o^{\rm T}(\bD^{++})^2 \omega \right\}\;.
\label{3.21}
\ee
Because of the identity $(\cD^{--})^2\cD^{++} q^+ =\cD^{++}(\cD^{--})^2 
q^+$,
the dynamical equation $\cD^{++} q^+=0$ implies $(\cD^{--})^2 q^+ = 0$.
Therefore, we have
$(D^+)^4 (\bD^{--})^2 q^+ = 0$ on the mass shell. 
Because of the identity $(\cD^{--})^2(\cD^{++})^2 \o =
(\cD^{++})^2(\cD^{--})^2 \o $, the free dynamical equation
$(\cD^{++})^2 \o =0$ implies 
$(D^+)^4 (\bD^{--})^2 \o = 0$. Therefore, the on-shell superfields
satisfy the equations
\bea
 \Box \cV^{++} &+& [{\bar W}_0,[W_0,\cV^{++}]] = 0 \label{3.22}\\
 \Box q^+  &+& {\bar W}_0 W_0 \;q^+ = 0 \label{3.23}\\
 \Box \o  &+& {\bar W}_0 W_0 \;\o = 0 \label{3.24}
\eea
which determine the masses of the superfields. Let us notice again that 
$\cV^{++}$ is also restricted by the requirement (\ref{gk}).

Now, we turn to the analysis of the case $W_0 \neq 0$ and 
$\o_0 \neq 0$. Here we  
split $\o = \o_0 + \o_{{\rm dynamical}}$ and skip the subscript 
``dynamical'' for readability.
The matter gauge 
transformation (\ref{19-1}) takes the form
\be
\d \o = -{\rm i} \l \o_0 - {\rm i} \l \o \qquad \l = \l^a T_a \;.
\label{3.25}
\ee
This situation exactly corresponds to the standard Higgs mechanism
where some scalar fields can be gauged away due to the presence of 
non-vanishing vacuum expectation values. In our case we can completely 
gauge 
away several $\o$-hypermultiplets. But then we stay with a number of
massive $V^{++}$-superfields whose masses no longer satisfy eq. 
(\ref{3.10})
and are greater than the central 
charge values. Therefore, the massive gauge superfields
now describe the long massive vector multiplets \cite{fay}.  

Upon the splitting $V^{++}\longrightarrow V_0^{++} + \cV^{++}$ and
$\o \longrightarrow \o_0 + \o$, the classical action takes the form
\be
S[V_0^{++} + \cV^{++}, q^+, \o_0 + \o] =
S_{(2)}[\cV^{++}, q^+,\o] + S_{\rm int}[\cV^{++}, q^+,\o]
\label{3.26}
\ee
where
\bea
&&S_{(2)}
=\frac{1}{2g^2}{\rm tr}\,\int d^{12}zdu_1du_2\,\frac
{\cV^{++}_\tau(1)\cV^{++}_\tau(2)}{(u^+_1u^+_2)^2}
-\frac{1}{2} \int d\z^{(-4)} du \,\o_0^{\rm T} (\cV^{++})^2\o_0\non\\
&& + \int d\z^{(-4)} du \,\left\{
-\breve{q}^+\bD^{++}q^+
+\frac{1}{2} \o^{\rm T}(\bD^{++})^2 \omega 
+{\rm i}\o^{\rm T}_0 \cV^{++} \bD^{++} \o
\right\}
\label{3.27}
\eea
and $S_{\rm int}$ includes the third- and higher-orders in the 
dynamical superfields. The $S_{\rm lin}$ is invariant
under the linearized gauge transformations
\be
\d \cV^{++} = \bD^{++}\l \qquad \d q^+ = 0 \qquad \d \o = - {\rm i} \l 
\o_0 \qquad \l = \l^a T_a
\label{3.28}
\ee
which can be used to impose some sort of unitary gauge on $\o$ just to
eliminate the mixed $\cV$--$\o$ term in the action.

Finally, in the third variant $W_0 = 0$, $\o_0 \neq 0$ we deal with 
$N=2$
supersymmetry without central charges in the spontaneously broken phase.
All the hypermultiplets remains massless, but there appear massive
gauge superfields which realize the first type of $N=2$ massive vector 
multiplets \cite{fay,fay2,gikos}. This picture has been described in
\cite{ddf}. 
 
Up to now we have discussed only mass generation by vacuum expectation 
values of scalar fields and have not considered explicit mass terms.
This does not restrict the validity of our discussion because
mass terms for hypermultiplets can be written as vacuum expectation 
values
of scalar fields. A mass term for hypermultiplets is 
equivalent to their coupling to a background abelian
gauge superfield $\G^{++}_0$ with constant strength \cite{cent}
\be
\G^{\pm \pm}_0 = -(\q^\pm)^2 {\bar w}_0 -({\bar \q}^\pm)^2  w_0 \qquad
 w_0 = \frac{1}{4} ({\bar D}^+)^2 \G^{--}_0\;.
\label{3.29}
\ee
Here $w_0$ is a fixed constant of mass dimension, and all information
about the masses of the hypermultiplets is encoded in the $U(1)$ 
generator
$\cM$, to which $\G^{++}_0$ is associated and which should commute
with the gauge group, $[\cM , T_a] = 0$. Then, the matter action 
reads
\bea
S_{{\rm MAT}}[V^{++}, q^+, \o]&=&
- \int d\z^{(-4)}
du \,\breve{q}^+ (\nabla^{++} + {\rm i}\G^{++}_0 \cM)q^+ \non\\
&+& \frac{1}{2}\int d\zeta^{(-4)}
du\,\omega^{\rm T}(\nabla^{++} + {\rm i}\G^{++}_0 \cM)^2\omega\;.
\label{3.30}
\eea
By construction, this theory possesses $N=2$ supersymmetry
with central charges. In eqs. (\ref{3.4}) and (\ref{3.5}) 
$\nabla^{++}$ is shifted to $\nabla^{++} + {\rm i}\G^{++}_0 \cM$. 
The analog of eq. (\ref{3.9}) reads
\be
[{\bar W}_0, W_0] = 0 \qquad (W_0 + w_0 \cM)\,\o_0 = 
({\bar W}_0 + {\bar w_0} \cM)\,\o_0 = 0\;.
\label{3.31}
\ee

\sect{ Effective equations of motion}

Long ago, West \cite{west} showed that the perturbative quantum 
corrections
can not remove the degeneracy in the classical vacuum solutions for
the supersymmetric theories with unbroken supersymmetry. Here we extend
this result to account for non-perturbative quantum corrections in the 
pure
$N=2$ supersymmetric gauge theory. The general form of 
low-energy effective action 
$\G[V^{++}]$, including non-perturbative quantum corrections, 
in the pure $N=2$ supersymmetric gauge theory reads 
\cite{sei,gat,wgr,pw,gru,cl,lgru} 
\bea
\G[V^{++}]&=& {\rm tr}\,\int d^4 x d^4\q F(W) + 
{\rm tr}\,\int d^4 x d^4{\bar \q} {\bar F}({\bar W})\non\\
&+& {\rm tr}\,\int d^4 x d^4 {\bar \q} d^4 \q H(W,{\bar W})
\label{4.1}
\eea
with holomorphic $F(W)$ and  Hermitian 
$H(W,{\bar W})$ functions
of the super field strengths. In the framework of perturbation theory, 
the holomorphic part of the effective action was found by Seiberg 
\cite{sei}
by integrating the anomaly of $R$-symmetry. The result was 
rederived it in terms of 
$N=1$ superfields \cite{wgr,pw} and $N=2$ superfields
\cite{max,back}. The non-perturbative holomorphic effective action
was found by Seiberg and Witten \cite{sw}. It has also been shown, using
the $N=1$ supergraph technique \cite{wgr,pw} and the $N=2$ harmonic
superspace approach \cite{max}, that the effective potential gets
perturbative non-holomorphic corrections to $H({\bar W}, W)$ .  

It is an instructive exercise to obtain the
effective equations of motion 
\be
\frac{\d \G[V^{++}]}{\d V^{++}}=0
\label{4.2}
\ee
As is shown in Appendix C, the variational derivative of 
$\G[V^{++}]$ is
\bea
\frac{\d \G[V^{++}]}{\d V^{++}}&=&
(\cD^+)^2 F'(W) + ({\bar \cD}^+)^2 {\bar F}'({\bar W})\non\\
&+&\frac{1}{16}(\cD^+)^2 ({\bar \cD}^+)^2
\left\{({\bar \cD}^-)^2 \frac{ \pa H(W,{\bar W})}{\pa W}
+ (\cD^-)^2 \frac{\pa H(W,{\bar W})}{\pa {\bar W}}\right\}  \;.
\label{4.3}
\eea
Each classical vacuum solution (\ref{5})
of the pure $N=2$ supersymmetric gauge theory satisfies the effective equations 
of motion (\ref{4.2}), so the vacuum expectation values are not changed by
quantum corrections. 

\vspace{1cm}

\noindent
{\bf Acknowledgements.}
We are grateful to Friedemann Brandt for fruitful discussions.
This work was supported by
the RFBR-DFG project No 96-02-00180,
the RFBR project No 96-02-16017 and by the Alexander
von Humboldt Foundation.

\begin{appendix}
\sect{Conventions}

We use the Lorentz and two-component spinor notations and conventions 
adopted
in \cite{des}. The $SU(2)_A$ indices are raised and lowered by $\ve^{ij}$
and $\ve_{ij}$, $\ve^{12}=\ve_{21}=1$, in the standard fashion
\be 
C^i = \ve^{ij}C_j \qquad C_i = \ve_{ij}C^j\;.
\ee
The $SU(2)$-invariant matrices
$(\t^I)_i{}^j \equiv \s^I$, where $I=1,2,3$, and their descendants
\bea
&(\t^I)_{ij}  \equiv  \ve_{jk} (\t^I)_i{}^k =(\t^I)_{ji} \qquad
(\t^I)^{ij}  \equiv  \ve^{ik} (\t^I)_k{}^j =(\t^I)^{ji} \non \\
&(\t^I)^{ij} (\t^I)_{kl} = -(\d^i_k \d^j_l + \d^i_l \d^j_k)
\eea
are used to convert a real triplet $D^I$ into the symmetric isotensor
\be
D^{ij} = (\t^I)^{ij} D^I \qquad \qquad D^I = -\frac{1}{2} (\t^I)_{ij} 
D^{ij}
\label{cr}
\ee
which satisfies the reality condition
\be
\overline{D^{ij}} = - D_{ij}\;.
\label{rc}
\ee

\sect{Bosonic action}

In this appendix we consider the bosonic sector of the general $N=2$
supersymmetric gauge theory (\ref{3.1}) in components. 
To pass to components, it is 
useful to choose the Wess-Zumino gauge (\ref{34}). It turns out that
only the leading (bosonic) components of $q^+$ and $\breve{q}^+$
\bea
q^+(x_A,\q^+,{\bar \q}^+,u) &=& u^{+i} C_i(x_A) + \cdots \non\\
\breve{q}^+(x_A,\q^+,{\bar \q}^+,u) &=& - u^+_i {\bar C}^i(x_A) + \cdots
\qquad {\bar C}^i \equiv \overline{C_i}
\label{b.1}
\eea
constitute the physical fields.  the remaining fields, denoted by dots,
are auxiliary. Similarly, 
the bosonic sector of $\o$ reads
\bea
&\o (x_A,\q^+,{\bar \q}^+,u) = A(x_A) + {\rm i}\, B^{ij}(x_A) u^+_i 
u^-_j + \cdots \non\\
&{\bar A} = A \qquad \qquad \overline{B^{ij}} = -B_{ij}
\label{b.2}
\eea
and the fields indicated by dots have no independent dynamics.

The Lagrangian of bosonic fields looks like
\bea 
\cL_{{\rm BOS}} &=& - \frac{1}{2g^2} {\rm tr}\,F^{mn} F_{mn}
+ \frac{1}{g^2} {\rm tr}\, \pa^m {\bar N} \pa_m N \non\\
&{}& + \nabla^m{\bar C}^i \nabla_m C_i + \nabla^m A^{{\rm T}} \nabla_m A
+ (\nabla^m B^I)^{{\rm T}} \nabla_m B^I \non\\
&{}& + \hat{D}^{Ia}\hat{D}^{Ia} - \cP(\varphi)
\label{b.3}
\eea
where
\bea
&\nabla_m = \pa_m + {\rm i}\,V_m\qquad
F_{mn} = \pa_m V_n - \pa_n V_m - {\rm i}\, [V_m,V_n]\label{b.4} \\
&\hat{D}^{Ia} = \frac{1}{3g} D^{Ia} + {\rm i}\,g A^{{\rm T}} T_aB^I
-\frac{1}{2}g {\bar C}^i (\t^I)_i{}^j T_a C_j\;.
\label{b.5}
\eea
The scalar potential reads
\bea
\cP(\varphi)&=& \frac{1}{4g^2} {\rm tr}\, \left( [{\bar N}, N]\right)^2
+ g^2 \left( \frac{1}{2} {\bar C}^i (\t^I)_i{}^j T_a C_j - 
{\rm i}\, A^{{\rm T}} T_a B^I \right)^2 \non\\
&{}& + {\bar C}^i \{{\bar N}, N\} C_i + \frac{1}{2} 
A^{{\rm T}} \{{\bar N}, N\} A + 
\frac{1}{2} (B^I)^{{\rm T}} \{{\bar N}, N\} B^I
\label{b.6}
\eea

\sect{Derivation of the effective equations of motion}

In this appendix we derive eq. (\ref{4.3}). We consider the effective 
action 
as a functional of the unconstrained analytic prepotential $V^{++}$, i.e.
 $W$ and $\bar W$ are given by (\ref{30}). If we make use of the
crucial relation 
\be
\nabla^{++}\d V^{--} = \nabla^{--} \d V^{++}\,,
\label{a1}
\ee
which follow from (\ref{26}) and (\ref{28}), we can determine the 
variation
of $V^{--}$ with respect to an arbitrary variation of $V^{++}$.
For simplicity, we will handle only the holomorphic functional
\be 
\cF =  {\rm tr}\,\int d^4 x d^4\q\, F(W)\;.
\ee
The other terms can be treated similarly.

We start with the identities
$$
\d \cF ={\rm tr}\int d^4 x d^4\q\,\d W F'(W)
={\rm tr}\int d^4 x d^4\q du\,\d W F'(W)
={\rm tr}\int d^4 x d^4\q du\,\d W_\l F'_\l(W_\l)
$$
and insert here the expression for $W_\l$ (\ref{30}). Next, the covariant
$u$-independence of $W_\l$, eq. (\ref{a1}) and the identities 
\be
({\bar \nabla}^+)^2 = [\nabla^{++},\nabla^{+}\nabla^{-}] \qquad
({\bar \nabla}^-)^2 = [\nabla^{--},\nabla^{+}\nabla^{-}]
\ee
allow us to continue as follows
\bea
\d \cF &=& \frac{1}{4}{\rm tr}\int d^4 x d^4\q du\, ({\bar \nabla}^+)^2
\d V^{--} F'(W_\l) 
=-\frac{1}{4}{\rm tr}\int d^4 x d^4\q du 
\,\nabla^{+}\nabla^{-}\nabla^{++}
\d V^{--} F'(W_\l) \non\\
&=&-\frac{1}{4}{\rm tr}\int d^4 x d^4\q 
du\,\nabla^{+}\nabla^{-}\nabla^{--}
\d V^{++} F'(W_\l)=\frac{1}{4}{\rm tr}\int d^4 x d^4\q du\, 
({\bar \nabla}^-)^2
\d V^{++} F'(W_\l)\;. \non
\eea
The next step is to formally rewrite
\be
F'(W_\l) = W_\l \frac{F'(W_\l)}{W_\l} =\frac{1}{4} ({\bar \nabla}^+)^2 
V^{--}
 \frac{F'(W_\l)}{W_\l}\;.
\ee
Then one gets
\bea
\d \cF &=& \frac{1}{16}{\rm tr}\int d^4 x d^4\q du\, ({\bar \nabla}^-)^2
({\bar \nabla}^+)^2 \left\{ \d V^{++} V^{--} 
\frac{F'(W_\l)}{W_\l}\right\} 
\non \\
&=& \frac{1}{16}{\rm tr}\int d^4 x d^4\q du\, ({\bar D}^-)^2
({\bar D}^+)^2 \left\{ \d V^{++} V^{--} \frac{F'(W_\l)}{W_\l}\right\}\non 
\\
&=&{\rm tr}\int d^{12} du \, \d V^{++} V^{--}
\frac{F'(W_\l)}{W_\l} \non \\
&=& \frac{1}{16}{\rm tr}\int d \z^{(-4)}du\, (\nabla^+)^2 
({\bar \nabla}^+)^2 \left\{ \d V^{++} V^{--}
\frac{F'(W_\l)}{W_\l}\right\}\;.
\eea
Since $W$ is covariantly chiral, we finally obtain
\be
\d \cF =\frac{1}{4}{\rm tr}\int d \z^{(-4)}du\, \d V^{++} (\nabla^+)^2 
F'(W_\l)\;.
\ee

\end{appendix}



\begin{thebibliography}{99}
\bibitem{sw} N. Seiberg and E. Witten, Nucl. Phys. B426 (1994) 19;
B430 (1994) 485.
\bibitem{bil} A. Bilal, {\it Duality in N=2 SUSY SU(2) Yang-Mills Theory:
A Pedagogical Introduction to the Work of Seiberg and Witten},
hep-th/9601007.
\bibitem{ler} W. Lerche, {\it Introduction to Seiberg-Witten Theory
and its Stringy Origin}, hep-th/9611190.
\bibitem{agh} L. Alvarez-Gaum\'e and S.F. Hassan, {\it Introduction to 
S-Duality in N=2 Supersymmetric Gauge Theories}, hep-th/9701069. 
\bibitem{n2} S. Ferrara and B. Zumino, Nucl. Phys. B79 (1974) 413;
P. Fayet, Nucl. Phys. B113 (1976) 135;
L. Brink, J.H. Schwarz and J. Scherk, Nucl. Phys. B121 (1977) 77.
\bibitem{gsw} R. Grimm, M. Sohnius and J. Wess, Nucl Phys. B133 (1978) 
275.
\bibitem{s} M. Sohnius, Nucl. Phys. B136 (1978) 461.
\bibitem{gikos} A. Galperin, E. Ivanov, S. Kalitzin, V. Ogievetsky and
E. Sokatchev, Class. Quantum Grav. 1 (1984) 469.
\bibitem{gios} A. Galperin, E. Ivanov, V. Ogievetsky and E. Sokatchev,
Class. Quantum Grav. 2 (1985) 601, 617.
\bibitem{zup} B.M. Zupnik, Teor. Mat. Fiz. 69 (1986) 207;
Phys. Lett. B183 (1987) 175.
\bibitem{max} I.L. Buchbinder, E.I. Buchbinder, E.A. Ivanov, S.M. 
Kuzenko and B.A. Ovrut, {\it Effective Action of the N=2 Maxwell Multiplet in 
Harmonic Superspace}, hep-th/9703147.
\bibitem{back} I.L. Buchbinder, E.I. Buchbinder, 
S.M. Kuzenko and B.A. Ovrut, {\it The Background-Field Method for N=2
Super Yang-Mills Theories in Harmonic Superspace}, hep-th/9704214.
\bibitem{des} N. Dragon, U. Ellwanger and M. Schmidt,
Prog. Particle Nucl. Phys. 18 (1987) 1.
\bibitem{ddf} B. Delamotte, F. Delduc and P. Fayet, Phys. Lett.
B176 (1986) 409.
\bibitem{fay2} P. Fayet, Nucl. Phys. B246 (1984) 89.
\bibitem{fay} P. Fayet, Nucl. Phys. B149 (1979) 137.
\bibitem{ohta} N. Ohta, Phys. Rev. D32 (1985) 1467; N. Ohta, H. Sugata
and H. Yamaguchi, Ann. Phys. 172 (1986) 26.
\bibitem{gio} A. Galperin, E. Ivanov and V. Ogievetsky, Nucl. Phys. 
B282 (1987) 74.
\bibitem{cent} S.M. Kuzenko, {\it The Off-Shell Massive Hypermultiplets 
Revisited}, hep-th/9704002.
\bibitem{west} P.C. West, Nucl. Phys. B106 (1976) 219.
\bibitem{sei} N. Seiberg, Phys. Lett. B206 (1988) 75.
\bibitem{gat} S.J. Gates, Nucl. Phys. B238 (1984) 349.
\bibitem{wgr} B. de Wit, M.T. Grisaru and M. Ro\v{c}ek, Phys. Lett. B374
(1996) 297.
\bibitem{pw} A. Pickering and P. West, Phys. Lett. B383 (1996) 54.
\bibitem{gru} M.T. Grisaru, M. Ro\v{c}ek and R. von Unge, Phys. Lett. B383
(1996) 415.
\bibitem{cl} T.E. Clark and S.T. Love, Phys. Lett. B388 (1996) 577.
\bibitem{lgru} U. Lindsr\"{o}m, F. Gonzales-Rey, M.Ro\v{c}ek and R. von
Unge, Phys. Lett. B388 (1996) 581.

\end{thebibliography}
\end{document}